\documentclass{optica-article}

\journal{opticajournal} 

\articletype{Research Article}

\usepackage{lineno}
\usepackage{multirow}
\usepackage{changes}
\usepackage{todonotes}
\usepackage{graphicx}
\usepackage{subcaption}
\usepackage{pdfpages}

\begin{document}

\title{Universal Design Path for Optomechanical Crystals and One-dimensional Photonic Crystals}

\author{Berke Demiralp\authormark{1}, Nien-Hsuan Lee\authormark{1,3}, Eva M. Weig\authormark{1,2,4,*} }

\address{\authormark{1}Department of Electrical Engineering, School of Computation,
Information and Technology, Technical University of Munich, 85748 Garching, Germany\\
\authormark{2}Munich Center for Quantum Science and Technology (MCQST), 80799 Munich, Germany\\
\authormark{3}present address: Luxembourg Institute of Science and Technology (LIST), 4362 Esch-sur-Alzette, Luxemburg(LIST)\\
\authormark{4}TUM Center for Quantum Engineering (ZQE), 85748 Garching, Germany }

\email{\authormark{*}eva.weig@tum.de}

\begin{abstract*}

We have shown a pattern that connects the refractive index, area and the cavity modes of the optomechanical crystals (OMCs) by the same order function.
By keeping the fundamental and second cavity modes within a range of $\pm 16$\,nm and $\pm 23$\,nm we have shown the link between the design area of the OMC and the refractive index of the material, by keeping the design area same we have shown the link between the refractive index and the cavity mode wavelength and by keeping the refractive index the same, we have shown the link between the cavity mode wavelength and the design area.
We have performed simulations for 2 different OMC designs and 10 different refractive indices (9 different materials) to prove the first two claims and we have performed both simulations and experiments on a 3C-SiC OMC, which resulted as $100$\,nm shift of the second cavity mode, to prove the last claim.
Our findings prove that a universal design for optomechanical crystals is possible, making the transition to different material bases easier to exploit their specific properties, suggesting a path to commercialize such devices for hybrid quantum technologies and having flexibility of tuning such devices for their own relative applications.     

\begin{figure}[ht!]
\centering\includegraphics[width=7cm]{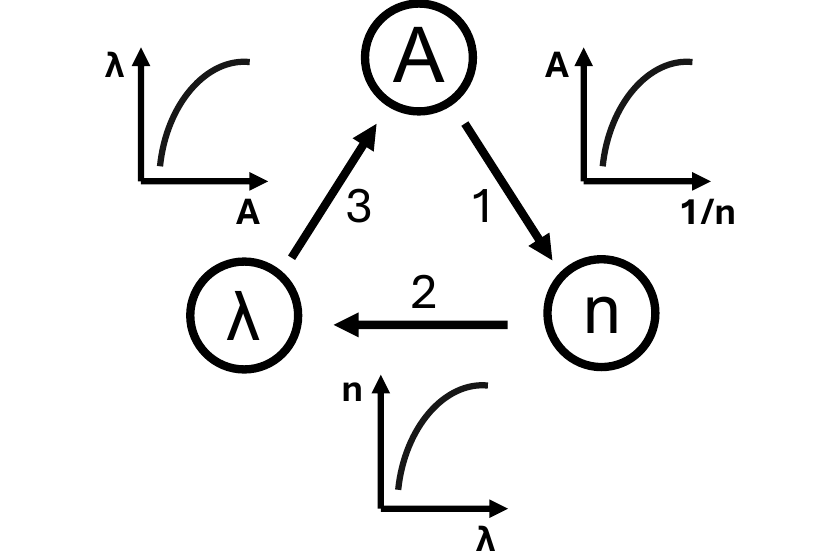}
\caption{Visualized representation of area, refractive index and cavity mode wavelength relation.
}
\label{fig:thecircle}
\end{figure}

\end{abstract*}

\section{Introduction}
Metamaterials, periodically structured materials, have attracted considerable interest in the last decades due to their unique and exotic applications, some of them being negative refractive index materials, lossless waveguide bends, polarization selective optics and selective broad band absorption/transmission \cite{liu2011metamaterials}. In particular, they allow one to engineer bandgaps for wave-like excitations such as light or sound, forming a photonic~\cite{krauss1999photonic} or phononic crystal~\cite{pennec2010two}, respectively. Their metamaterial-nature allows to control by means of their design how the underlying excitations can propagate. This is not only exploited to realize waveguides guiding those excitations, but can also be used to create localized defects by breaking the symmetry of the periodic structure to introduce discrete states in the bandgap. The resulting confined modes are referred to as photonic or phononic crystal cavities~\cite{yablonovitch1994photonic,lu2009phononic}. Owing to the periodicities achievable in nanostructured materials, combined with the respective propagation velocities of light and sound, the photonic resonances occur in the near-infrared range of the electromagnetic spectrum (approx. $200$\,THz) and the phononic resonances at $<10$\,GHz frequencies, respectively.

In recent years, metamaterials have been conceived that exhibit both photonic and phononic bandgaps and which allow for the co-localization of photonic and phononic cavity modes~\cite{maldovan2006simultaneous}. They allow for the study of dispersive light-sound interations, and represent an important platform for cavity-optomechanics~\cite{aspelmeyer2014cavity}. Experimental realizations of these so-called optomechanical crystals (OMCs)~\cite{eichenfield2009optomechanical} (or phoXonic cavities~\cite{oudich2014optomechanic,gomis2014one}
are usually based on a freely suspended quasi-one dimensional metamaterial slab. 

It is useful to classify OMCs according to their optical linewidth $\kappa$ with respect to the mechanical eigenfrequency $\Omega_\mathrm{m}$, defining the sideband-resolution of the system. In a sideband-resolved optomechanical system, $\kappa \ll \Omega_\mathrm{m}$ is achieved, whereas $\kappa \gg \Omega_\mathrm{m}$ holds for the non-resolved situation.

Optomechanical crystals have been employed to explore a variety of cavity-optomechanical phenomena. While hallmark effects such as the optical spring effect (optomechanical spring softening/hardening)~\cite{safavinaeini2011emit} or optomechanical cooling/heating~\cite{chan2011laser} in the classical regime of large phonon occupation numbers do not necessitate sideband-resolution, the situation changes when targeting the quantum regime where the phonon number approaches zero.

Operating in the resolved sideband regime gives access to the canonical optomechanical interaction, described by the linearized optomechanical Hamiltonian in the sideband‑resolved regime, yielding a beamsplitter‑type or two‑mode squeezing-type interaction for red or blue detuning of the laser with respect to the cavity mode, respectively~\cite{aspelmeyer2014cavity}. Entering this regime is not only a prerequisite for ground-state cooling~\cite{chan2011laser, Meenehan2015,Wang2023} and further quantum operations such as single phonon interferometry, 
remote mechanical entanglement, or the realization of a mechanical quantum memory~\cite{Hong2017,Riedinger2018, Wallucks2020}, but also for phonon routing or for creating photon-phonon correlations in the classical regime~\cite{Fang2016,Fang2017}.

All of the above-mentioned experimental demonstrations have been realized in silicon, which represents the standard material platform for OMCs. Within this platform, geometric variations have been developed to address specific applications, including OMCs with acoustic radiation shields to reduce phonon leakage~\cite{chan2012laser}, unreleased OMC designs which avoid the suspended geometry to improve robustness and thermal anchoring~\cite{kolvik2023clamped}, and quasi-two-dimensional OMC geometries offering simultaneous photonic and phononic bandgaps in a planar geometry, strong optomechanical coupling, 
and improved thermalization~\cite{Alegre2011, Safavi-Naeini2014, Ren2020, Mayor2025}.

Beyond silicon, a broad range of material platforms has been explored in photonic as well as optomechanical crystals to leverage material-specific properties not available in silicon. Examples include two-photon absorption-free silicon nitride~\cite{Davanco2014}, aluminum nitride offering an electrical link to mechanical and optical degrees of freedom via piezoelectric actuation~\cite{fan2013aluminum}, lithium niobate as a platform for mechanical coupling to microwaves~\cite{jiang2019lithium} and for high mechanical resonance frequency tuning~\cite{liang2017high}, gallium phosphide for mechanical lasing~\cite{schneider2019optomechanics}, microwave-to-optical conversion~\cite{honl2022microwave} and quantum memory applications~\cite{tamaki2024two}, diamond~\cite{burek2016diamond, Cady2019, Joe2024, Oh2025} as well as 3C and 4H silicon carbide~\cite{lu2020silicon,lee2015high,radulaski2013photonic,lukin20204h} for nonlinear mechanical and optical interactions and spin integration toward defect-based hybrid qubit systems interfacing optical and mechanical degrees of freedom~\cite{koppenhofer2023single}, yttrium iron garnet~\cite{rashedi2024yig} as a mechanical-magnetic interface and as a route toward connecting magnomechanics and optomechanics~\cite{engelhardt2022optimal}, and erbium-doped silicon for linear Stark tuning of optical transitions~\cite{yu2023frequency} and Purcell enhancement exploiting the high-$Q$ optical cavity~\cite{gritsch2023purcell}. The wide spectrum of applicability across different material platforms and physical figures of merit has already established OMCs as a promising building block for hybrid quantum technologies, as well as a versatile tool for studying fundamental physics~\cite{chu2020perspective,barzanjeh2022optomechanics}.

The broad variety of materials, geometries, and applications has resulted 
in a large number of independently developed OMC layouts. Although the 
design rules governing the defect region follow common principles --- 
namely, gradually increasing the material volume to elevate the effective 
refractive index and thereby localize the optical mode, with mode tuning 
achieved by modifying the defect density~\cite{chan2011laser, 
oudich2014optomechanic, lee2015high} --- no unified design framework 
applicable across different material platforms has been established, nor 
has a general relationship between mode wavelength and refractive index 
been proposed. Here, we introduce a framework for adapting OMC designs 
to a variety of materials, and propose a relationship between the defect 
region area $A$, the refractive index $n$, and the cavity mode wavelength 
$\lambda$, connecting these three quantities through a set of second-order 
polynomial functions, as illustrated in Fig.~\ref{fig:thecircle}. These 
relations enable scaling of an OMC layout between different material 
platforms without the need for manual re-optimization. The validity of 
the proposed framework is demonstrated using the first two optical cavity 
modes of two distinct OMC layouts, successfully recovering the target 
cavity modes across ten different material platforms, providing a 
systematic and transferable approach for translating OMC designs to 
novel materials.

In Section~2, we introduce the design parameters of the two OMC layouts 
considered in this work. Sections~3--5 focus on the optical properties of the OMC, and each address one leg of the 
diagram in Fig.~\ref{fig:thecircle} using finite element simulations using COMSOL Multiphysics complemented with experimental data. Section~3 is devoted to leg 1. We perform simulations 
in which the fundamental optical mode (and second optical mode) is 
maintained near $1550$\,nm ($1640$\,nm), with a maximum deviation of 
$16$\,nm ($23$\,nm), across two layouts and ten different refractive 
indices, by uniformly scaling the total OMC area. Section~4 discusses leg 2. With the 
OMC area held fixed, we simulate the relationship between refractive index 
and cavity mode wavelength. The 
Section~5 describes leg 3. At fixed refractive index, we simulate the resonance wavelength 
shift of both optical cavity modes as a function of area scaling; using 
one of the layouts implemented in thin-film 3C-SiC, we experimentally 
demonstrate this mode shift and use it to extract the refractive index of 
our material by cross-referencing with finite element simulations. Across all three legs, the relations between 
area, refractive index, and cavity mode wavelength are found to follow 
second-order polynomial functions.

For the mechanical analysis, the breathing mode was selected as the 
representative mode of interest due to its broad applicability. 
Simulations were performed for the same nine materials, revealing a linear 
relationship between the breathing mode frequency and the defect area 
(see Supplement~1 Fig. S4).

In Section~6, we assess the effect of area scaling on the optomechanical 
coupling rate and optical quality factors, benchmarking the framework 
against previously reported silicon and diamond OMC devices.

 \section{Design parameters}
 
\begin{figure}[ht!]
    \centering\includegraphics[width=7cm]{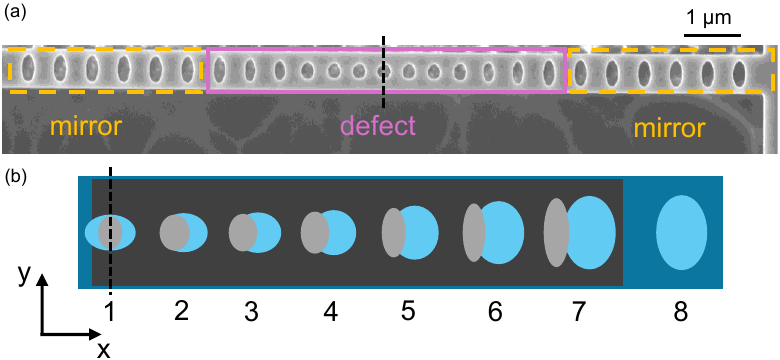}
    \caption{(a) Scanning electron microscope (SEM) image of fabricated D2, with mirror regions in the orange dashed rectangles and the defect region in the purple rectangle. Dashed black line divides the OMC into two symmetric parts from the center defect unit where each unit are numbered from 1 to 7. (b) Schematic illustration of the two investigated designs of the half of the defect region, D1 (blue) and D2 (gray). The black dashed line visualizes the symmetry axis in the center of the OMC. } 
    \label{fig:D1_D2}
\end{figure}

An OMC is realized in a rectangular nanobeam perforated with a periodic arrangement of elliptical holes, as illustrated by the scanning electron micrograph in Fig.~\ref{fig:D1_D2}(a). The arrangement of the holes defines three distinct sections: mirror regions on either side and a central defect region. The mirror regions consist of a periodic pattern of identical ellipses, realizing bandgaps for selected symmetry classes of photonic and phononic modes. The defect region is characterized by a smooth, adiabatic variation of the hole geometry toward the center, introducing discrete localized modes into each bandgap. This adiabatic tapering approach has been shown to suppress scattering losses and yield higher optical and mechanical quality factors compared to abrupt defect designs~\cite{Akahane2003,eichenfield2009optomechanical,chan2011laser}.

In this work, we focus on the defect region, as it is the part of the OMC where mode localization occurs. For a given beam thickness, the 
defect region geometry is parametrized by the beam width, the number of ellipses in the defect region, their horizontal and vertical diameters (width and height, respectively), and their center-to-center separation, commonly referred to as the unit cell length. 
We introduce an \textit{area multiplication factor} (AMF) that scales the beam width, ellipse width, ellipse height, and unit cell length by a single, uniform factor, while keeping the number of ellipses in the defect region constant. 
The mirror regions primarily determine the center frequency and width of the of the bandgap and do not significantly impact the defect mode; their geometry is therefore scaled alongside the defect region without further re-optimization. Notice that this procedure may influence the location of the defect mode within the band gap. As long as the defect mode remains far from the edges of the band gap, this does not significantly affect the performance of the device.%

Two OMC designs are studied in this work. They are denoted D1 and D2. D1 follows the design rules for the silicon platform, inspired by one of the earliest OMC designs featuring elliptical holes~\cite{chan2011laser}. Its geometry is illustrated in Fig.~\ref{fig:D1_D2}(b) in blue. 
In the defect region of D1, the minor axes of the ellipses along the $x$ direction follows a sinusoidal function, the major axes along the $y$ direction is described by a Gaussian function, whereas the center-to-center separation of the ellipses follow a linear function, all varying from the defect center to the mirror-defect boundary.
These functional forms were chosen to ensure a smooth, adiabatic transition of the mode profile from the mirror to the defect center (see Fig.~S2 of Supplement~1), minimizing scattering losses. Tables~S1, S2 and S3 of Supplement~1 list material properties and the exact geometry of the defect region of D1 as well as the mirror region. 
The beam width of D1 is $534$\,nm and the defect region comprises 15 unit cells. 
In silicon, D1 yields the fundamental and second optical modes at $1546$\,nm 
and $1633$\,nm, respectively, and a mechanical breathing mode at 
$5.16$\,GHz. 

D2 was developed in-house for the 3C-SiC platform and is illustrated in Fig.~\ref{fig:D1_D2}(b) in gray. 
As in D1, the minor axes of the ellipses follows a Gaussian function and their center-to-center separation follows a linear function, but with different parameter values. The major axes of the ellipses follows a sinusoidal function, resulting in minor ellipse axes approximately half those of D1. 
Tables~S1, S2 and~S3 of Supplement~1 also include material properties and the exact geometry of the defect region of D2 as well as the mirror region. The beam width of D2 is $500$\,nm and the defect region comprises 13 unit cells.
In silicon carbide, D2 yields the fundamental and second optical modes at $1551$\,nm and $1637$\,nm, respectively, and a mechanical breathing mode at $6.49$\,GHz. 

For both designs, Fig.~\ref{fig:D1_D2}(b) shows the central unit cell (labelled as cell 1) and the right half of the defect region. Material properties and the exact defect as well as mirror region dimensions of both designs are listed in Tables~S1, S2 and~S3 of Supplement~1, respectively. The silicon realization of D1 serves as the reference from which scaling to other material platforms is performed in Sections~3 and 4. In Section~5, D2 is implemented in 3C-SiC for both simulations and experiment. A beam thickness of $220$\,nm is assumed for all materials in Sections~3 and 4. In Section~5, the experimentally investigated devices are made from 3C-SiC-on-Si with a film thickness of $210$\,nm. The $10$\,nm difference in thickness introduces a constant shift in the resonance frequencies of approx. $1\%$, but does not affect the scaling relations between defect area, refractive index, and cavity mode wavelength derived in this work.

The materials used throughout the simulations and their refractive indices are listed in Table~S1 of Supplement~1. In all simulations, a 
fixed refractive index evaluated at $1550$\,nm is used for each material, i.e. material dispersion is neglected. This choice is 
deliberate: by fixing the refractive index, we eliminate the influence of wavelength-dependent dispersion from our results and demonstrate that the scaling relations derived in this work arise purely from the scaled geometry, rather than from the wavelength dependence of the refractive index. This approximation is well justified within our framework, which spans a refractive index range of $2.00$ to $3.48$: across all investigated materials, the maximum dispersion-induced refractive index variation over the simulated wavelength range of $954$\,nm to $1650$\,nm is $\Delta n = 0.127$, observed for GaAs~\cite{polyanskiy2024refractiveindex}. 
This represents a small perturbation relative to the refractive index differences between platforms that the framework is designed to bridge.

\section{Area and refractive index relation}

Our starting point is the inverse relationship between the refractive index of a material $n_\text{mat}$ and the wavelength of the mode 
confined within it: to compensate for an increase (decrease) in refractive index, the area available to the mode must be decreased 
(increased) accordingly. This relation forms the first leg of the diagram in Fig.~\ref{fig:thecircle}, connecting the refractive index 
axis to the area axis. 
We define the area multiplication factor (AMF) as the uniform factor by which all in-plane geometric parameters of the 
defect region (except the number of defect unit cells) are scaled relative to the reference silicon design, while the beam thickness is 
held fixed. The goal of this section is to establish a functional relationship between refractive index ratio $n_\text{Si}/n_\text{mat}$ and the AMF required to retain the fundamental optical mode at $1550$\,nm in the target material.

\begin{figure}[ht!]
\centering\includegraphics[width=7cm]{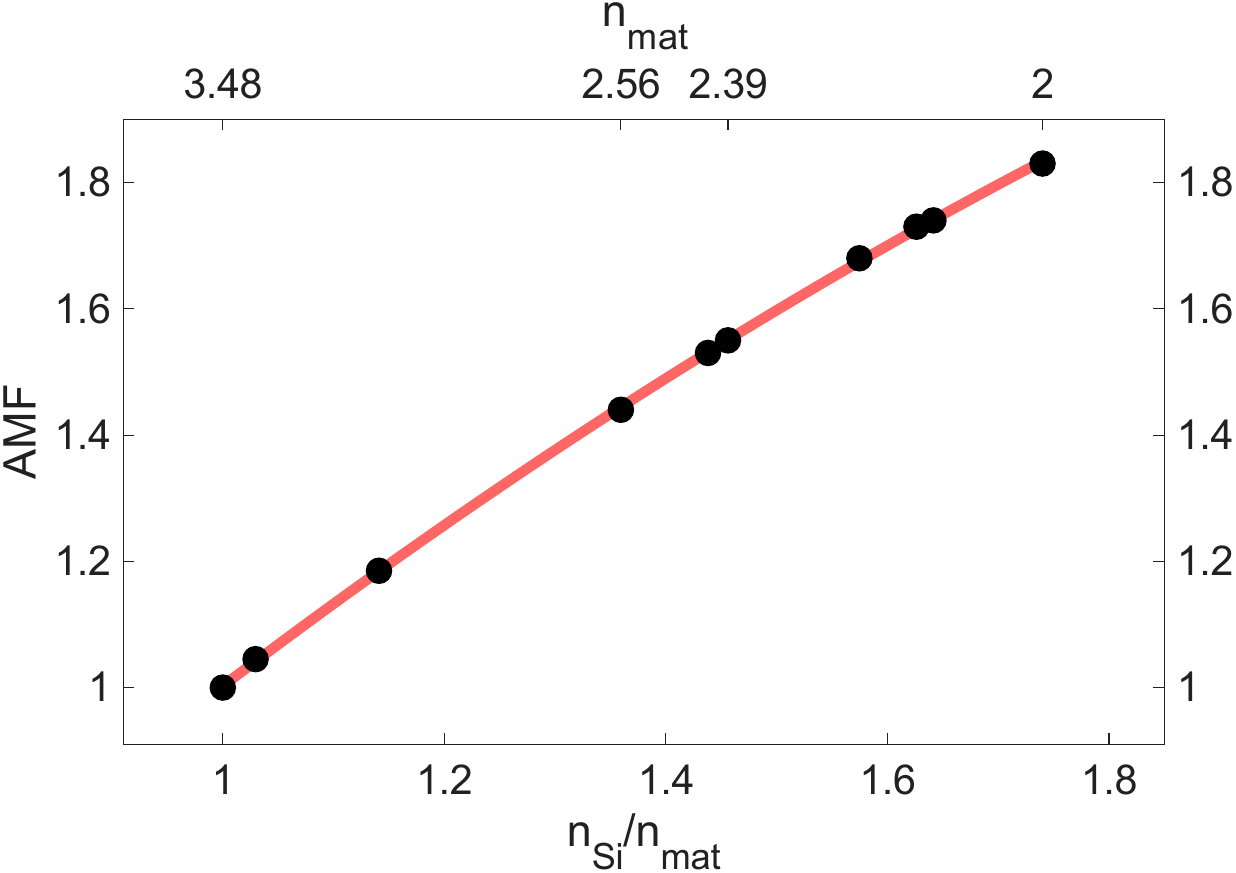}
\caption{AMF as a function of the refractive index of the $10$ target materials, c.f. Tab.~S1 of Supplement~1. Bottom (top) axis shows the refractive index ratio with respect to silicon (refractive index of target material). Black dots encoding the AMF coincide for the fundamental mode of D1 as well as D2, keeping the mode's wavelength within a tolerance of  $1550$\,\textpm $16$\,nm reflecting fabrication constraints. The same holds true for the second optical mode of D1 as well as D2, within a tolerance of $1640$\,\textpm $23$\,nm. A second-order polynomial is fitted to the data (red line).}
\label{fig:refindex_amf}
\end{figure}

To this end, we proceed as follows: For each of the ten material 
platforms specified in Tab.~S1 of Supplement~1, we use the refractive index ratio $n_\text{Si}/n_\text{mat}$ 
as a physically motivated initial estimate of the AMF and compute the 
resulting fundamental optical mode frequency. The AMF is then 
iteratively adjusted in steps of approximately $2\%$ until the 
fundamental optical mode falls within $1550 \pm 16$\,nm,
corresponding to a wavelength range of $1534$\,nm - $1566$\,nm. This tolerance is 
motivated by fabrication constraints: targeting $1550$\,nm exactly 
would require AMF adjustments at the sub-nanometer level, which is not 
meaningful in practice. The final value of the AMF at convergence is 
taken as the AMF for that material. 
Repeating this procedure across the ten materials yields a set of $(n_\text{mat}, \text{AMF})$ data 
points for each design, plotted in Fig.~\ref{fig:refindex_amf}. 
Notably, the data points obtained for D1 and D2 coincide within the 
simulation accuracy, indicating that the AMF-refractive index relation 
is independent of the specific OMC layout. This observation motivates 
fitting a single phenomenological function to the combined dataset, from 
which the functional relationship between $n_\text{mat}$ and the AMF 
can be identified.

\begin{figure}[ht!]
\centering\includegraphics[width=7cm]{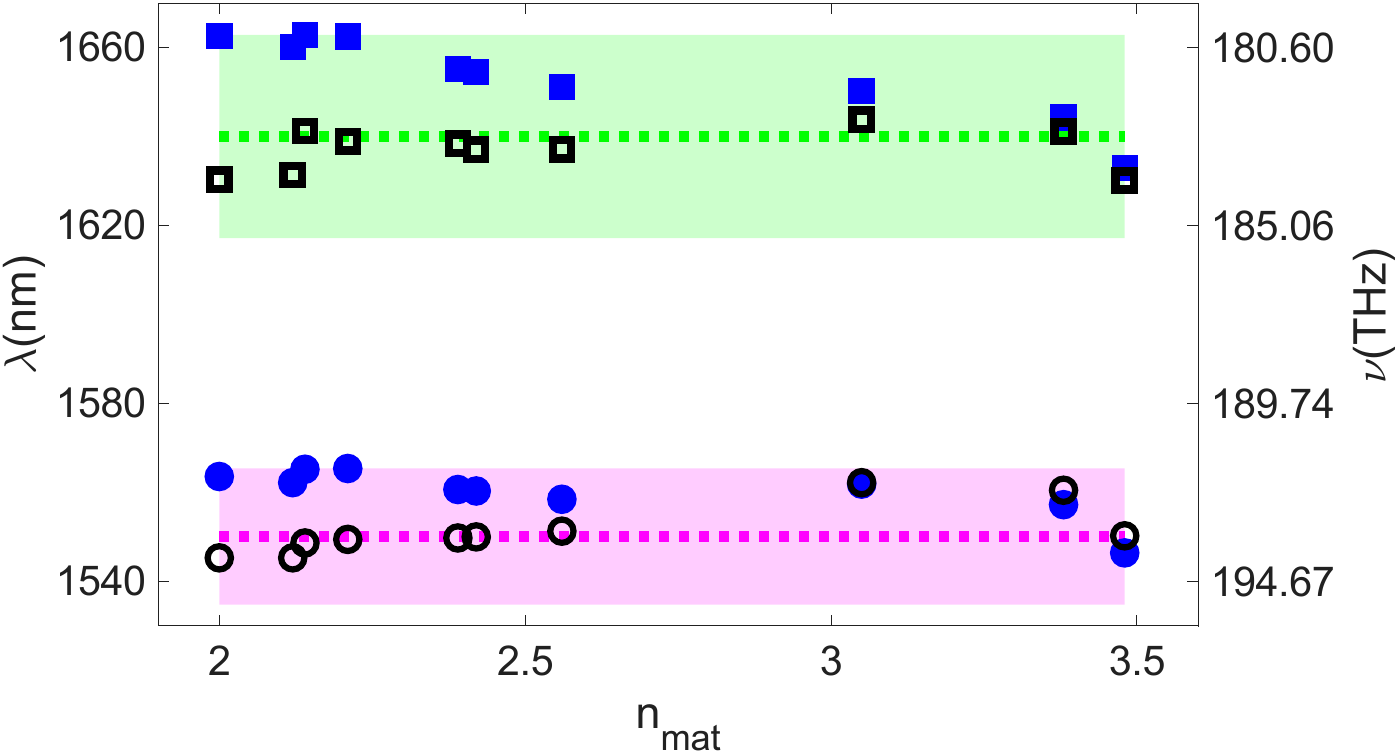}
\caption{Wavelengh $\lambda$ and frequency $\nu$ of the fundamental and second optical cavity modes of D1 (solid blue circle and square, respectively) and D2 (hollow black circle and square, respectively). Dotted pink (green) line illustrates the fundamental (second) optical mode target wavelength of $1550$\,nm ($1640$\,nm).}
\label{fig:1st_2nd_line}
\end{figure}

Fitting a second-order polynomial to the combined $(n_\text{Si}/n_\text{mat}, 
\text{AMF})$ dataset yields
\begin{equation}
    \text{AMF}(n_\text{Si}/n_\text{mat}) = a \, (n_\text{Si}/n_\text{mat})^2 + b \, (n_\text{Si}/n_\text{mat}) + c,
      \label{eq:AMF_n}
\end{equation}
with coefficients $a = -0.2784$, $b = 1.886$, 
and $c = -0.6055$, valid for both D1 and D2 
(Fig.~\ref{fig:refindex_amf}). This second-order dependence, rather 
than the simple linear ratio $n_\text{Si}/n_\text{mat}$, reflects the 
fact that the effective mode confinement depends nonlinearly on the 
refractive index contrast.
The same second-order relationship is found for the second optical mode, 
second mode is maintained within $\pm 23$\,nm of $1640$\,nm, 
corresponding to a range of $1617$\,nm - $1663$\,nm.
Being able to scale both the fundamental and the second optical mode is particularly relevant when applying schemes that involve two different modes, e.g. for sideband thermometry~\cite{Safavi-Naeini2012}.
It is important to note that achieving the fundamental mode at exactly 
$1550$\,nm for a given design and material does not guarantee that the 
second mode will fall at exactly $1640$\,nm, and vice versa. This is illustrated in Fig.~\ref{fig:1st_2nd_line}, showing the exact wavelengths and frequencies of the fundamental and second optical mode for both D1 and D2 explicitly. The relative deviations of both modes 
from their respective target wavelengths, and their alternating 
character between D1 and D2, reflect the different defect geometries of 
the two designs: since the shape and volume of the defect region differ 
between D1 and D2, the two modes respond differently to the same AMF.

\section{Refractive index and cavity mode wavelength relation}

\begin{figure}[ht!]
\centering\includegraphics[width=7cm]{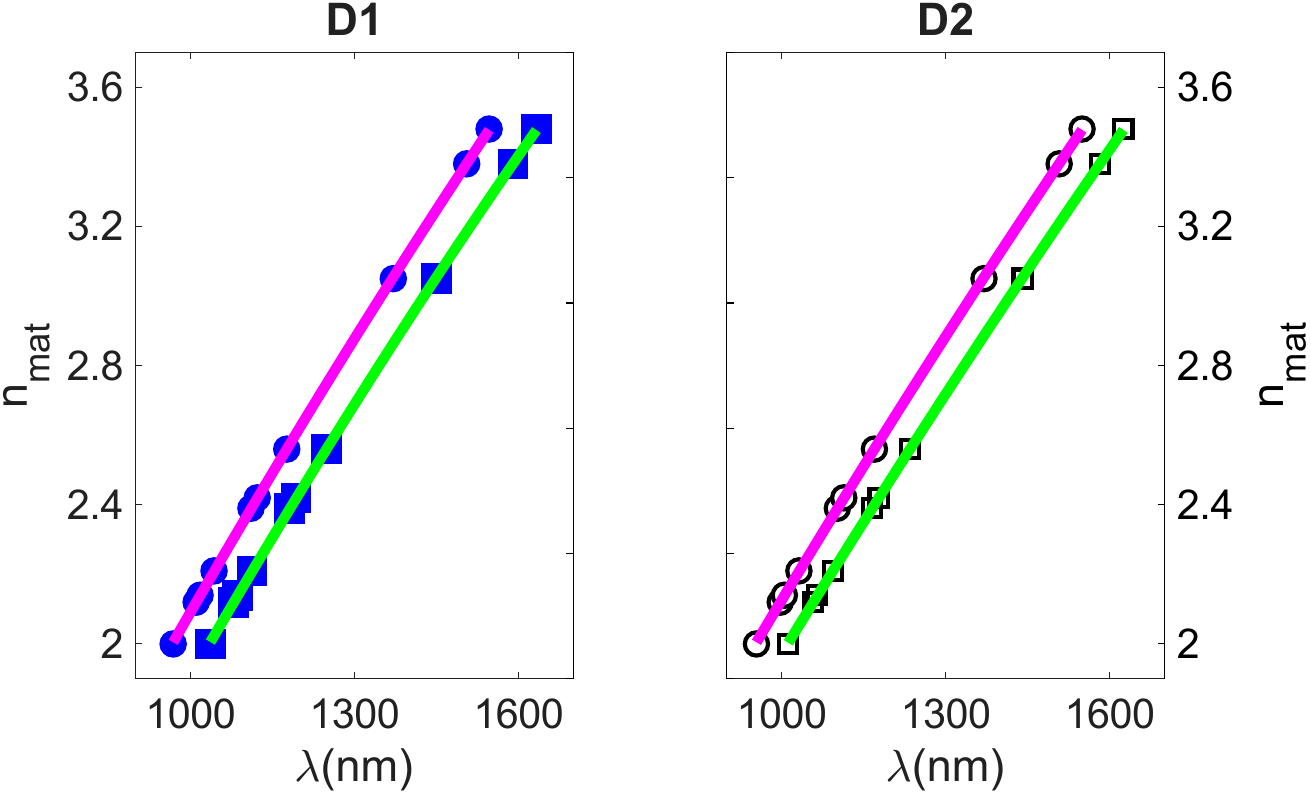}
\caption{Refractive index of target material plotted against the cavity mode wavelength of the fundamental and second optical cavity modes of D1 (left, solid blue circle and square, respectively) and D2 (right, hollow black circle and square, respectively) for fixed AMF $= 1$.
The pink (green) lines display the (shifted) second-order polynomial fitted to the fundamental (second) optical mode of D1 and D2.
The average wavelength separation between the two modes 
is $74$\,nm for D1 and $65$\,nm for D2.}
\label{fig:mode_wavelength}
\end{figure}

The second leg of the diagram in Fig.~\ref{fig:thecircle} connects the 
refractive index axis to the cavity mode wavelength axis, at fixed 
area. Keeping the AMF fixed at unity for both D1 and D2 --- i.e. 
retaining the reference silicon geometry without rescaling --- we 
computed the fundamental and second optical cavity mode wavelengths 
across the ten material platforms by varying only the refractive index, as depicted in Fig.~\ref{fig:mode_wavelength}.

For both modes and both designs, the cavity mode wavelength follows a 
second-order polynomial function of the refractive index, see Sec.~3 of Supplement~1.
While the functional form is the same 
for D1 and D2, the absolute mode wavelengths and their mutual spacing 
differ between the two designs, as a direct consequence of their 
different defect geometries and hence different effective refractive 
indices experienced by the confined mode. Specifically, the average 
wavelength separation between the fundamental and second modes is 
$74$\,nm for D1 and $65$\,nm for D2. The fact that a second-order 
polynomial describes this relation for both designs and both modes, 
despite these differences, further corroborates the universality of the 
functional relationships identified in Section~3.

\section{Area and cavity mode wavelength relation}

The third leg of the diagram in Fig.~\ref{fig:thecircle} connects the 
area axis to the cavity mode wavelength axis, at fixed refractive index. 
By keeping the material fixed and varying the AMF, the cavity mode 
wavelengths shift accordingly. This is explored in an experiment with OMCs using the 3C-SiC platform and design D2 
(see Fig.~\ref{fig:D1_D2}(a)).

\begin{figure}[ht!]
\centering\includegraphics[width=7cm]{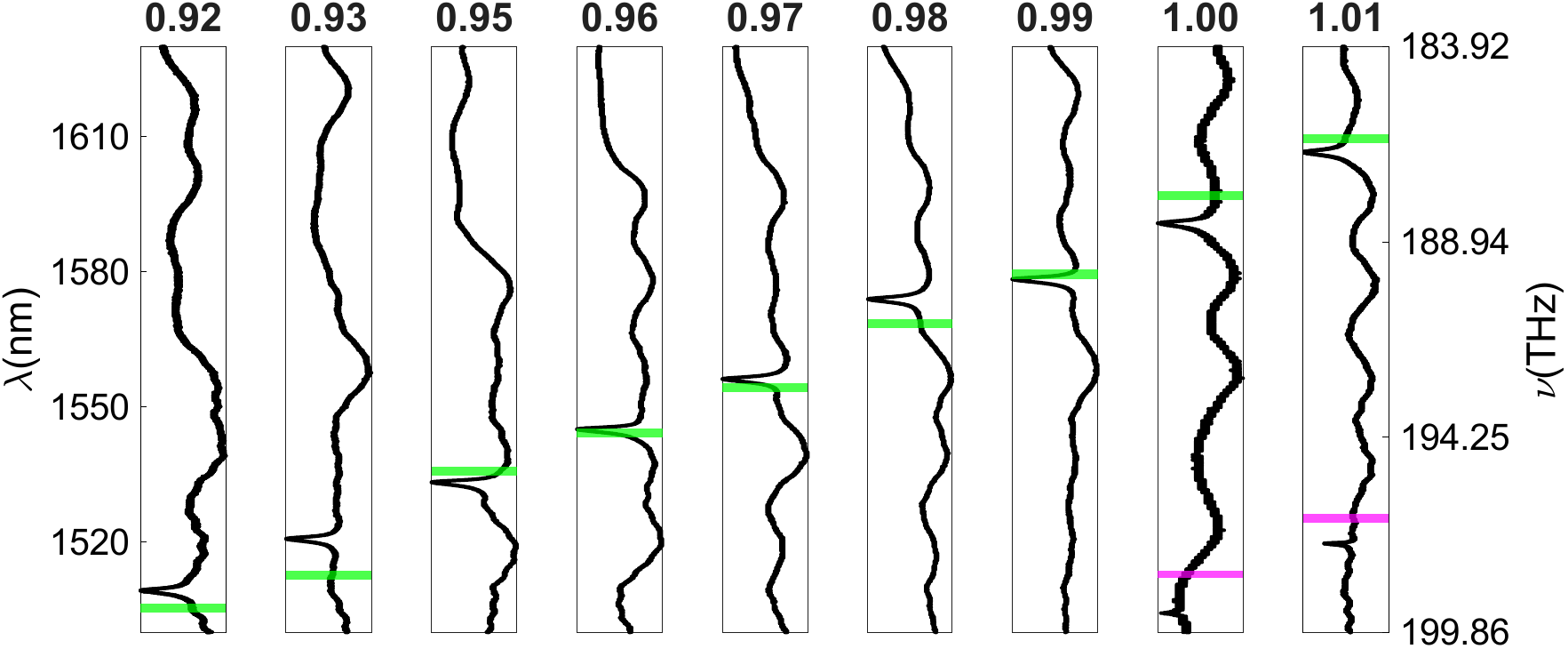}
\caption{Experimentally obtained spectra of 3C-SiC OMCs based on D2 for AMF increasing from $0.92$ to $1$ in steps of $0.01$. The AMF value obtained from SEM analysis of the device geometry is indicated above each dataset; slight deviation from nominal AMF results from fabrication imperfection.
Pink and green bars specify the wavelength of the fundamental and second optical cavity mode simulated with a refractive index $n = 2.52$ at $1550$\,nm. 
The fundamental cavity mode is only observed in the available frequency tuning range for AMFs $\ge 1$, while the second mode is observed in all devices.}
\label{fig:experiment_refIndex}
\end{figure}

A set of devices was fabricated with AMF values decreasing from $1$, i.e. the D2 layout specified in Tabs.~S2 and~S3 of Supplement~1, in steps of 
$0.01$ to $0.8$. The fundamental and second optical cavity modes were 
measured in the wavelength range $1500$\,nm - $1630$\,nm, defined by the 
tuning range of the laser used in the experiment. Figure~\ref{fig:experiment_refIndex} shows the obtained spectra that yield modes in that wavelength range (see Sec.~5 of Supplement~1 for experimental setup). For increasing AMF, we observe a continuous redshift of both cavity 
modes across the measurement window: the second mode shifts across the 
entire accessible wavelength range, while the fundamental mode is 
observed over a partial range of $1500$\,nm - $1520$\,nm, consistent with 
its higher-frequency origin. 
Simulations were performed using the same D2 geometry and a beam 
thickness of $210$\,nm, matching the experimentally characterized 3C-SiC 
film thickness. The simulated wavelengths of the fundamental and second cavity modes are included in Fig.~\ref{fig:experiment_refIndex} as pink and green bars, respectively.

Comparing the simulated and measured mode wavelengths, we find that the 
refractive index of our thin-film 3C-SiC is $n = 2.52$ at $1550$\,nm, 
extracted by taking the second mode as the reference. This value differs 
from the literature value of $n = 2.56$ (see Tab.~S1 of Supplement~1), which does not reproduce the measured cavity mode wavelengths.
We attribute the discrepancy to a combination of fabrication-induced material modifications 
and an insufficient quality of the literature value, resulting from a limited availability of optical characterization data for 
thin-film 3C-SiC at near-infrared wavelengths.

\begin{figure}[ht!]
\centering\includegraphics[width=7cm]{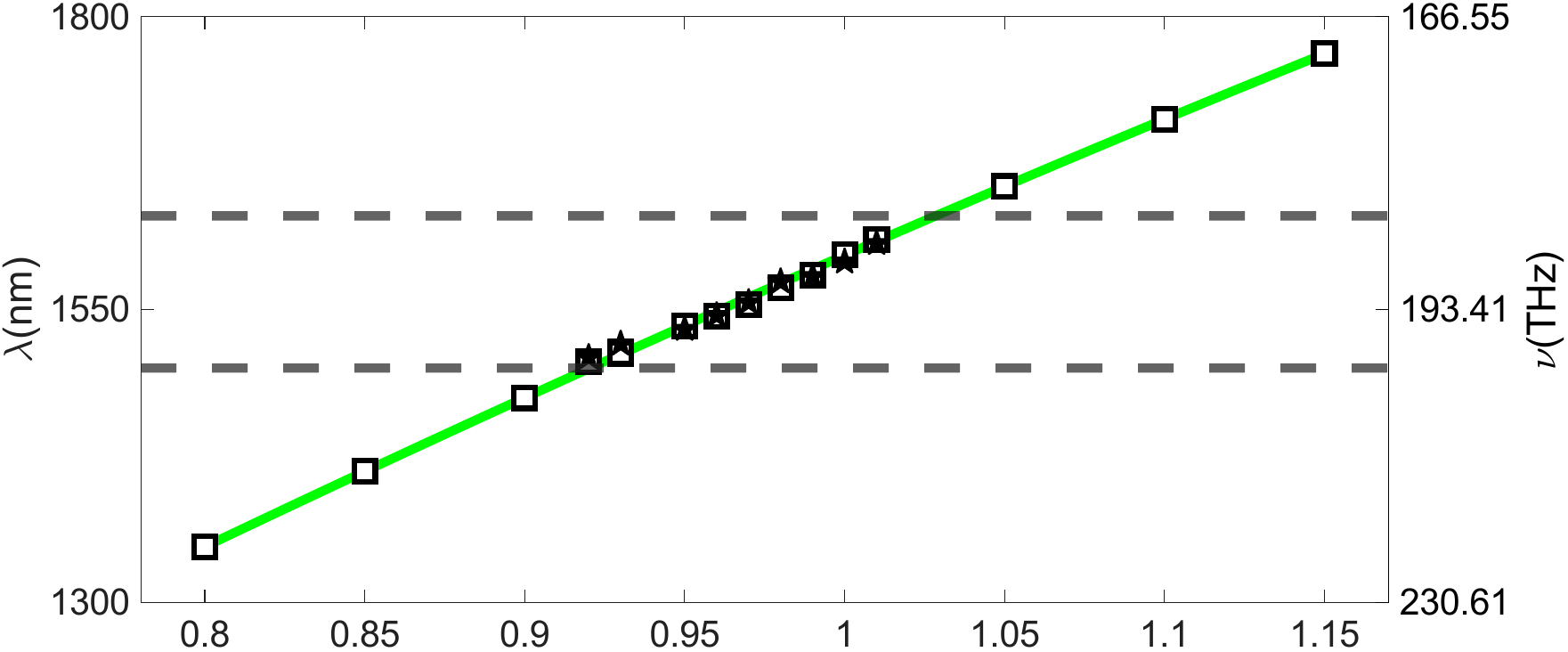}
\caption{Wavelength of the second optical cavity mode as a function of AMF for 3C-SiC 
OMCs based on D2, with refractive index $n = 2.52$. Experimental values from Fig.~\ref{fig:experiment_refIndex} are included as black crosses. Simulated values over the full AMF range ($0.80$ - $1.15$ in steps of $0.05$, along with those from Fig.~\ref{fig:experiment_refIndex} are included as hollow black squares. Polynomial relation is depicted as a green solid line. Dashed lines indicate the tuning range of the laser used 
in the experiment ($1500$\,nm - $1630$\,nm).}
\label{fig:experiment_AMF_largeScan}
\end{figure}

The wavelengths of the second optical cavity mode extracted from Fig.~\ref{fig:experiment_refIndex} are plotted as black crosses as a function of AMF in Fig.~\ref{fig:experiment_AMF_largeScan}. Simulated values using $n = 2.52$ 
for the 3C-SiC refractive index are included for AMF values between $0.80$ and $1.15$ in steps of $0.05$ as black open squares, going beyond the experimentally accessible range between the dashed lines. The simulated values matching the experimentally explored AMF covering the range of $0.92$ to $1.01$ in steps of $0.01$ from Fig.~\ref{fig:experiment_refIndex} are also included.

\section{Impact of scaling on optomechanical figures of merit}

To verify that the scaling approach does not compromise the other 
critical figures of merit of OMCs, we perform finite element simulations of the optical quality factor, the mechanical mode eigenfrequency as well as quality factor, and the vacuum optomechanical coupling rate for D1 and D2 scaled to silicon, diamond, and 3C-SiC. This allows to compare the novel 3C-SiC platform with the well-established OMC platform, silicon, and with the most closely related material, diamond. 
Both designs D1 and D2 are scaled to target the fundamental optical mode at 
$1550$\,nm as described in Section~3 in all three materials, and the resulting figures of merit are extracted from 
the simulations. The results are summarized in 
Tab.~\ref{tab:Q+g0}. 

\begin{table}[ht!]
\centering
\begin{tabular}{llccccc}
\hline
\textbf{Material} & \textbf{Design} & $\nu$ (THz) & $Q_\mathrm{opt}$ & 
$\Omega_\mathrm{m}$ (GHz) & $Q_\mathrm{m}$ & $|g_0|/2\pi$ (kHz) \\
\hline
\multirow{2}{*}{Silicon} 
    & D1 & $193.87$ & $563{,}000$ & $5.15$ & $2{,}660{,}000$ & $1{,}083$ \\
    & D2 & $193.48$ & $436{,}000$ & $7.05$ & $159{,}000$ & $1{,}239$ \\
\hline
\multirow{2}{*}{Diamond} 
    & D1 & $192.1$ & $27{,}000$  & $6.59$ & $4{,}100{,}000$ & $410$ \\
    & D2 & $193.49$ & $34{,}000$  & $9.02$ & $64{,}000$ & $409$ \\
\hline
\multirow{2}{*}{3C-SiC} 
    & D1 & $192.37$ & $43{,}000$ & $4.74$ & $3{,}430{,}000$ & $218$ \\
    & D2 & $193.31$ & $62{,}000$ & $6.49$ & $92{,}000$ & $472$ \\
\hline
\end{tabular}
\caption{Simulated optical eigenfrequency of the fundamental mode $\nu$, optical quality factor $Q_\mathrm{opt}$, mechanical eigenfrequency $\Omega_\mathrm{m}$, mechanical quality factor $Q_\mathrm{m}$, and vacuum 
optomechanical coupling rate $|g_0|/(2\pi)$ for D1 and D2 scaled to 
silicon, diamond, and 3C-SiC.}
\label{tab:Q+g0}
\end{table}

Importantly, the scaling procedure results in eigenfrequencies of the mechanical breathing mode $\Omega_\mathrm{m}$ in the physically accessible, single-digit GHz range across all investigated platforms, confirming the mechanical viability of the scaling approach.
See Sec.~4 of Supplement~1 for more details on other materials.
Notice that the simulated mechanical quality factor yields unphysically high results, as the simulation typically does not include the dominant damping mechanisms related to material and/or surface defects \cite{Cady2019}.

For 
silicon and diamond, where OMC implementations have been reported, we 
benchmark our results against the literature~\cite{chan2012optimized, 
Cady2019}. 
For silicon, our results are in good agreement with the simulations reported in~\cite{chan2012optimized}, reproducing the remarkably high optical quality factors and coupling rates found in this material platform. 
For diamond, the obtained values are also in the same range as those found in the literature~\cite{Cady2019}.
This confirms that our scaling procedure does not compromise the optomechanical figures of merit. 

For 3C-SiC, no literature reference on OMCs exists to date to the best of our knowledge, and 
our results therefore represent the first estimates of these figures of merit for this material platform.
The optical quality factor exceeds that of diamond, while the optomechanical vacuum coupling is comparable to that found in diamond for both designs. While D1 yields a high $Q_\mathrm{opt}$ at the expense of a reduced $|g_0|/(2\pi)$, D2 yields  significantly higher $Q_\mathrm{opt}$ as well as $|g_0|/(2\pi)$. Keeping in mind that the product $Q_\mathrm{opt} \times |g_0|$ is a common figure of merit for the optimization of OMCs, D1 is approximately on par with diamond while D2 surpasses the performance of the diamond platform. 

Finally, we note that neither D1 nor D2 are optimized for any of the specific figures of merit discussed in Tab.~\ref{tab:Q+g0}, but wish to point out that the mirror geometry in particular offers a straightforward route to 
improving $Q_\mathrm{opt}$ and $|g_0|/2\pi$ beyond the values reported 
here. This is, however, beyond the scope of this work.

\section{Conclusion}

We have introduced a scaling framework that connects the defect region 
area, the refractive index, and the optical cavity mode wavelength of 
an optomechanical crystal through a set of second-order polynomial 
relations. The framework is validated using two distinct defect region 
designs, D1 and D2, across ten material platforms spanning a refractive 
index range of $2.00$ to $3.48$, and for two optical cavity modes.

The three legs of the framework were established as follows. The 
relation between the area multiplication factor (AMF) and the refractive 
index was obtained by simulating the fundamental optical cavity mode 
across ten materials, recovering it within $1550 \pm 16$\,nm for both 
designs using a single second-order polynomial function. The same 
functional form is found for the second optical cavity mode, recovered 
within $1640 \pm 23$\,nm. The relation between cavity mode wavelength 
and refractive index at fixed area was established by simulation across 
the same ten platforms. Finally, the relation between cavity mode 
wavelength and area at fixed refractive index was explored both in 
simulation, covering a wavelength range of $954$\,-$1650$\,nm, and 
experimentally in 3C-SiC OMCs based on D2, covering the range 
$1500$\,-$1630$\,nm. The experimental results are in good agreement 
with simulations and additionally yield a refractive index of 
$n = 2.52$ at $1550$\,nm for our thin-film 3C-SiC, differing from the 
literature value of $n = 2.56$.

The consistency of the second-order polynomial relations across two 
designs, ten refractive indices, and two cavity modes demonstrates that 
the framework is robust and design-independent. This establishes a 
universal scaling path for OMCs and one-dimensional photonic crystals 
that is independent of both the material platform and the specific 
defect region design. New material platforms can be addressed by simply 
applying the AMF derived from the refractive index relation, without 
manual re-optimization of the device geometry. This significantly 
reduces the design effort required to translate an OMC to a new 
material, facilitating both mass fabrication and the integration of 
OMCs into hybrid quantum systems that exploit the unique properties of 
specific material platforms.

\begin{backmatter}

\bmsection{Funding}
We gratefully acknowledge financial support from the Deutsche Forschungsgemeinschaft (DFG, German Research Foundation) under Germany's Excellence Strategy, EXC-2111-390814868. This research is part of the Munich Quantum Valley, which is supported by the Bavarian State Government with funds from the Hightech Agenda Bayern Plus.

\bmsection{Acknowledgment}
We would like to thank Andreas Reiserer and his group for providing the tapered fiber and for valuable discussions on the experimental setup. The authors would also like to thank Simon Gröblacher for valuable discussions on the experimental setup and the optomechanical coupling simulations.   

\bmsection{Disclosures}
The authors declare no conflicts of interest.

\bmsection{Data Availability Statement}
Data underlying the results presented in this paper are available in Ref.~\cite{data}.

\bmsection{Supplemental Document} 
See Supplement~1 for supporting content.

\end{backmatter}

\bibliography{sample}



\section*{Supplementary material}

# Supplementary Material: Universal Design Path for Optomechanical Crystals and One-dimensional Photonic Crystals

This supplemental document includes detailed information about (1) material parameters, (2) the simulated optomechanical crystal design geometries, as well as (3) the obtained polynomial scaling relations. It further specifies (4) further optomechanical figures of merit, and describes (5) the experimental measurement setup.

## 1. MATERIAL PARAMETERS

Table S1 lists the values of the refractive index of the employed materials at 1550 $nm$ which have been employed in the simulations.

| Material | Symbol | Refractive index |
|---|---|---|
| Silicon Nitride | $Si_3N_4$ | 2 |
| Aluminum Nitride | AlN | 2.12 |
| Lithium Niobate n(e)/n(o) | $LiNbO_3$ | 2.14/2.21 |
| Diamond | C | 2.39 |
| Yttrium Iron Garnet | $Y_3Fe_2(FeO_4)_3$ | 2.42 |
| Silicon Carbide (3C/4H) | SiC | 2.56 |
| Gallium Phosphide | GaP | 3.05 |
| Gallium Arsenide | GaAs | 3.38 |
| Silicon | Si | 3.48 |

**Table S1.** Refractive index values of the materials under investigation at 1550 nm from the refractive index database *https://refractiveindex.info/* [1]

## 2. OPTOMECHANICAL CRYSTAL DESIGNS

Figure S1 provides a more detailed version of Fig. 2 of the main text. The SEM micrograph of Fig. S1(a) introducing the defect (mirror) region by purple (orange) boxes is repeated from Fig. 2(a) for the sake of convenience and clarity. The black dashed line indicates the symmetry line and marks the central ellipse located at position $(0, 0)$. The right half of the OMC is schematically shown in more detail in Fig. S1(b), detailing the geometry of both explored designs, D1 (top, in blue) and D2 (bottom, in gray). The black dashed line indicates the symmetry line and marks the central ellipse numbered as 1. The right half of the defect region is highlighted by a purple solid line. It comprises ellipses 1 to 8 for design D1 (blue) and 1 to 7 for design D2 (gray), respectively. The right mirror region is identified by a yellow dashed line. It comprises six ellipses labelled 9 to 14 for both for D1 and D2 (blue and gray, respectively; ellipse number 8 is omitted in D2).

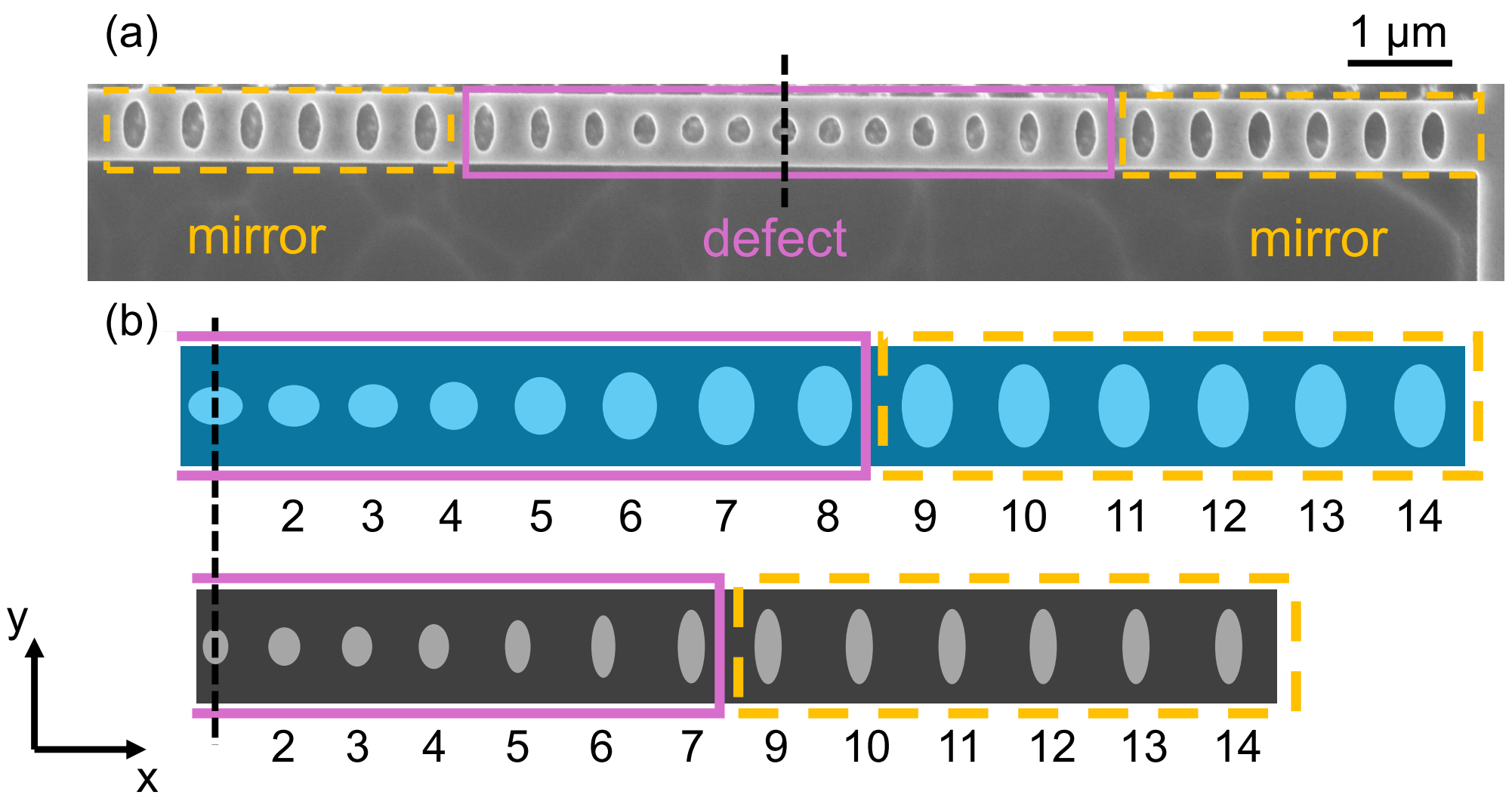


**Fig. S1.** (a) Scanning electron microscope (SEM) image of fabricated D2, with mirror regions in the orange dashed rectangles and the defect region in the purple rectangle. Dashed black line divides the OMC into two symmetric parts from the center defect located at $(0, 0)$. (b) Schematic illustration of the two investigated designs for D1 (top, blue) and D2 (bottom, gray). The black dashed line visualizes the symmetry axis in the center of the OMC. The right half of the defect region includes ellipses 1 to 8 (7) for D1 (D2), marked by a purple solid line. The right mirror region includes ellipses 9 to 14 for both designs, marked by a yellow dashed line.

### A. Defect region dimensions for D1/D2

Table S2 shows the diameters of the ellipses along the x- and y-direction in the defect region and their distance from the center of the OMC for D1 and D2 discussed in Section 2 of the main text (see Fig. S1 for details).

| Ellipse number | x diameter (nm) | y diameter (nm) | center position (nm) |
|---|---|---|---|
| 1 | 242.2/113.3 | 171.7/160.7 | 0/0 |
| 2 | 229/143.5 | 184.9/173.0 | 341.3/319.3 |
| 3 | 220.2/138.0 | 191.6/179.2 | 687/642.8 |
| 4 | 215.7/135.2 | 215.7/201.9 | 1048.1/980.6 |
| 5 | 229/114.6 | 253.2/236.9 | 1433.4/1341.1 |
| 6 | 242.2/110.4 | 295/280.2 | 1838.5/1720.2 |
| 7 | 246.6/124.2 | 345.7/323.4 | 2263.4/2117.8 |
| 8 | 242.2/N/A | 352.3/N/A | 2699.4/N/A |

**Table S2.** Defect region dimensions for D1/D2 using the labelling conventions introduced in Fig. 2 of the main text and Fig. S1.

Figure S2 depicts the adiabatic functions producing a smooth transition of the minor/major ellipses axis as well as the ellipse position from the mirror region.

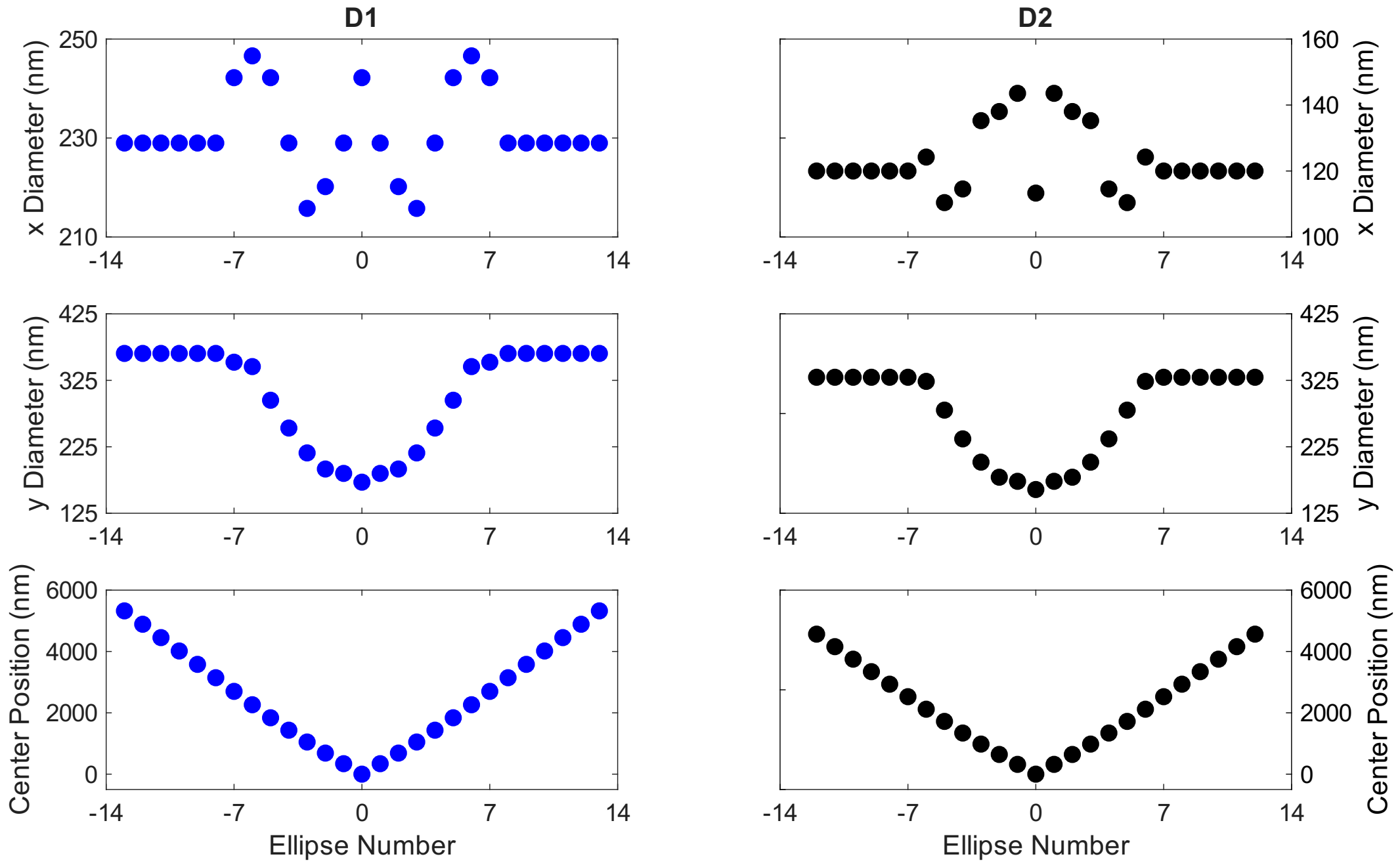


**Fig. S2.** Functional forms diameters of the ellipses along the x- and y-direction in the defect region and their distance from the center of the OMC of D1 (left, blue) and D2 (right, black), ensuring a smooth, adiabatic transition of the mode profile from the mirror to the defect center.

### B. Mirror region dimensions for D1/D2

Table S3 shows the diameters of the ellipses along the x- and y-direction in the mirror region and their distance from the center of the OMC for both designs D1 and D2 discussed in Section 2 of the main text (see Fig. S1 for details).

| Ellipse number | x diameter (nm) | y diameter (nm) | center position (nm) |
|---|---|---|---|
| 9 | 229/120 | 365.5/329.6 | 3144.2/2525.7 |
| 10 | 229/120 | 365.5/329.6 | 3580.1/2933.6 |
| 11 | 229/120 | 365.5/329.6 | 4016.1/3341.5 |
| 12 | 229/120 | 365.5/329.6 | 4452/3749.4 |
| 13 | 229/120 | 365.5/329.6 | 4888/4157.3 |
| 14 | 229/120 | 365.5/329.6 | 5324/4565.2 |

**Table S3.** Mirror region dimensions for D1/D2, continuing the ellipse count introduced in Fig. 2 of the main text, c.f. Fig. S1.

### C. AMF values for D1 and resulting fundamental and second mode

Table S4 shows the AMF values and exact simulated frequencies for design D1 discussed in Section 3 of the main text.

| Material | Refractive index | AMF | First Mode (THz) | Second Mode (THz) |
|---|---|---|---|---|
| Silicon Nitride | 2 | 1.83 | 191.74 | 180.32 |
| Aluminum Nitride | 2.12 | 1.74 | 191.91 | 180.59 |
| Lithium Niobate n(e)/n(o) | 2.14/2.21 | 1.73/1.68 | 191.54/191.52 | 180.29/180.33 |
| Diamond | 2.39 | 1.55 | 192.1 | 181.12 |
| Yttrium Iron Garnet | 2.42 | 1.53 | 192.14 | 181.2 |
| Silicon Carbide (3C/4H) | 2.56 | 1.44 | 192.37 | 181.57 |
| Gallium Phosphide | 3.05 | 1.185 | 191.95 | 181.67 |
| Gallium Arsenide | 3.38 | 1.045 | 192.52 | 182.32 |
| Silicon | 3.48 | 1 | 193.87 | 183.6 |

**Table S4.** AMF values for D1 and resulting eigenfrequency of the fundamental as well as second optical cavity mode.

### D. AMF values for D2 and resulting fundamental and second mode frequencies.

Table S5 shows the AMF values and exact simulated frequencies for design D2 discussed in Section 3 of the main text.

| Material | Refractive index | AMF | First Mode (THz) | Second Mode (THz) |
|---|---|---|---|---|
| Silicon Nitride | 2 | 1.83 | 194.01 | 183.89 |
| Aluminum Nitride | 2.12 | 1.74 | 194.01 | 183.77 |
| Lithium Niobate n(e)/n(o) | 2.14/2.21 | 1.73/1.68 | 193.59/193.49 | 182.66/182.92 |
| Diamond | 2.39 | 1.55 | 193.45 | 182.98 |
| Yttrium Iron Garnet | 2.42 | 1.53 | 193.42 | 183.13 |
| Silicon Carbide (3C/4H) | 2.56 | 1.44 | 193.26 | 183.12 |
| Gallium Phosphide | 3.05 | 1.185 | 191.91 | 182.38 |
| Gallium Arsenide | 3.38 | 1.045 | 192.12 | 182.68 |
| Silicon | 3.48 | 1 | 193.39 | 183.91 |

**Table S5.** AMF values for D2 and resulting eigenfrequency of the fundamental as well as second optical cavity mode.

## 3. POLYNOMIAL SCALING RELATIONS

### A. Scaling relations for the three legs

Figure 1 of the main text, reproduced here as Fig. S3, illustrates the relations between the area of the defect region, the refractive index of the material, and the cavity mode wavelength in a triangle.

In the following, we detail the second-order polynomial functions which have been obtained as scaling function for the first, second and third leg of the triangle,

$$\mathrm{AMF} = a_1\ (n_{\mathrm{Si}}/n_{\mathrm{mat}})^2 + b_1\, n_{\mathrm{Si}}/n_{\mathrm{mat}} + c_1, \tag{S1}$$

$$n = a_2\,\lambda^2 + b_2\,\lambda + c_2, \tag{S2}$$

$$\lambda = a_3\,\mathrm{AMF}^2 + b_3\,\mathrm{AMF} + c_3, \tag{S3}$$

with coefficients $a_i$, $b_i$ and $c_i$ for $i = 1, 2, 3$ listed in Table S6.

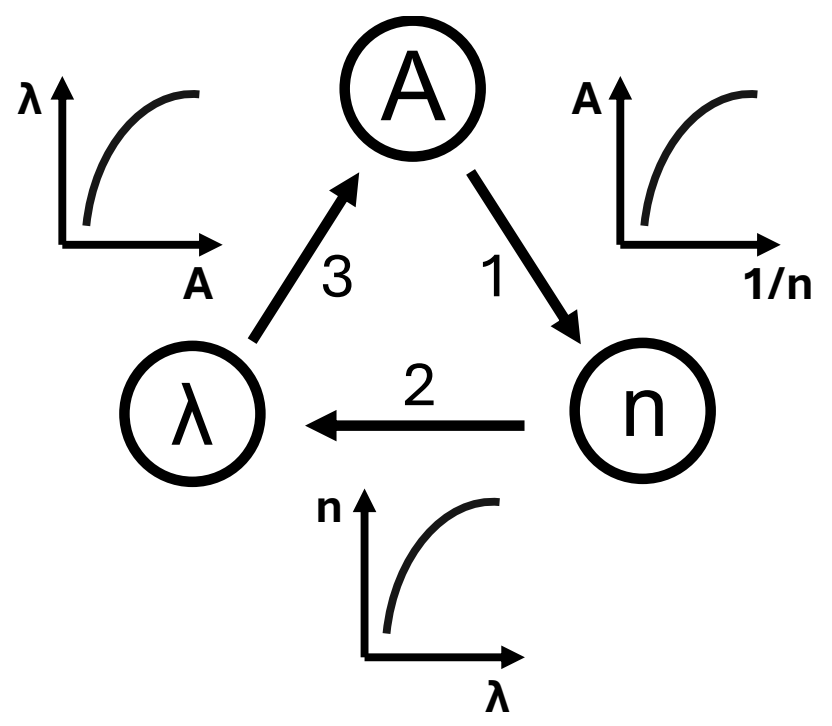


**Fig. S3.** Visualized representation of area, refractive index and cavity mode wavelength relation.

| Leg | Coefficient | D1 | D2 |
|---|---|---|---|
| | $a_1$ | -0.2784 | same as D1 |
| 1 | $b_1$ | 1.886 | same as D1 |
| | $c_1$ | -0.6055 | same as D1 |
| | $a_2$ | -3.362 [-3.751] x $10^{-7}$ | -2.663 [-2.587] x $10^{-7}$ |
| 2 | $b_2$ | 3.395 [3.47] x $10^{-3}$ | 3.146 [3.087] x $10^{-3}$ |
| | $c_2$ | -0.9684 [-1.189] | -0.7568 [-0.8668] |
| | $a_3$ | N/A | -279.3 |
| 3 | $b_3$ | N/A | 1747 |
| | $c_3$ | N/A | 128.3 |

**Table S6.** Coefficients of the second-order polynomials constituting the scaling functions of the three legs of the triangle in Fig. S3. Leg 1: Geometric scaling coefficients connecting defect region area - refractive index coincide for both designs. Leg 2: refractive index - wavelength scaling coefficients of the fundamental [second] optical cavity mode vary slightly between design D1 and D2. Leg 3: Wavelength - defect region area scaling coefficients have only been experimentally determined for design D2.

### B. Equivalence of the polynomials for leg 2

The second leg of Fig. S3 of the main text describes the relation between the resonance wavelength and the refractive index. We consider the fundamental and the second optical mode for two different designs D1 and D2, which all occur at a different resonance wavelength. Thus this gives rise to four polynomial fits, for each of the explored materials.

However, those four fits result from the same second-order polynomial from which they are shifted to match the wavelength of the mode (see Fig. 5 of main text). In the following, we show that the obtained polynomials can be traced back to such a common origin. To this end, we first shift each of the four polynomials such that their wavelength coincides at the lowest refractive index. As a next step, we compute the relative root mean square errors across the overlapping wavelength range. The results are presented in Tab. S7.

All errors are of order 1% and thus within the set tolerances of the simulation, motivated by fabrication constraints as described in the main text.

Notice that the same argument could be made for the third leg. However, we only fabricated samples for design D2, hence no polynomial fit is available for a comparison.

## 4. OPTOMECHANICAL FIGURES OF MERIT

Besides the optical cavity eigenfrequency, which is at the center of the present manuscript, the main text provides a discussion of all other optomechanical figures of merit for three distinct

| Design-Mode | D1-Fundamental | D2-Second |
|---|---|---|
| D2-Fundamental | 1% | 0.8% |
| D1-Second | 0.7% | 1.4% |

**Table S7.** Root-mean-square error of the shifted and subtracted polynomial fits of leg 2 from Fig. S3 of the main text.

material platforms, comparing the novel 3C-SiC platform with the well-established OMC platform, silicon, and with the most closely related material, diamond, see Tab. 1 of the main text.

In the following, we extend this discussion to the other materials explored in this work. As our focus lies on the defect region geometry, we specifically address those quantities which are dominated by the design of the defect region of the OMC.

### A. Breathing mode frequencies for D1 and D2

Tables S8 and S9 show the mechanical eigenfrequency of the breathing mode $\Omega_m$ simulated for D1 and D2 across ten material platforms, for two cases. In the first case, labelled as "unscaled", the reference silicon geometry is applied without rescaling (AMF = 1), such that only the refractive index varies between materials while the device dimensions remain fixed. In the second case, labelled "scaled", the material-specific AMF derived from the second-order polynomial relation of Section 3 is applied, rescaling the in-plane geometry to recover the fundamental optical mode near 1550 nm. As expected, the breathing mode frequency changes significantly upon rescaling, since the mechanical frequency scales inversely with the device dimensions. However, the shift is predictable and directly proportional to the AMF (see Fig. S4), such that all scaled designs remain within the single-digit GHz range across all investigated platforms, confirming the mechanical viability of the scaling approach.

| Material | Breathing Mode $\Omega_m$ (GHz) unscaled (AMF = 1) | AMF | Breathing Mode $\Omega_m$ (GHz) scaled (AMF as indicated) |
|---|---|---|---|
| Silicon Nitride | 5.39 | 1.83 | 2.95 |
| Aluminum Nitride | 6.19 | 1.74 | 3.56 |
| Lithium Niobate n(e)/n(o) | 3.64/3.64 | 1.73/1.68 | 2.10/2.17 |
| Diamond | 10.24 | 1.55 | 6.60 |
| Yttrium Iron Garnet | 3.76 | 1.53 | 2.46 |
| Silicon Carbide (3C/4H) | 6.84 | 1.44 | 4.75 |
| Gallium Phosphide | 8.83 | 1.185 | 7.42 |
| Gallium Arsenide | 2.43 | 1.045 | 2.32 |
| Silicon | 5.16 | 1 | 5.16 |

**Table S8.** Eigenfrequency of the breathing mode for D1 for unscaled geometry and scaled geometry with indicated AMF.

| Material | Breathing Mode $\Omega_m$ (GHz) unscaled (AMF = 1) | AMF | Breathing Mode $\Omega_m$ (GHz) scaled (AMF as indicated) |
|---|---|---|---|
| Silicon Nitride | 7.36 | 1.83 | 4.03 |
| Aluminum Nitride | 8.46 | 1.74 | 4.88 |
| Lithium Niobate n(e)/n(o) | 4.97/4.97 | 1.73/1.68 | 2.88/2.97 |
| Diamond | 13.99 | 1.55 | 9.03 |
| Yttrium Iron Garnet | 5.13 | 1.53 | 3.37 |
| Silicon Carbide (3C/4H) | 9.34 | 1.44 | 6.49 |
| Gallium Phosphide | 12.06 | 1.185 | 10.20 |
| Gallium Arsenide | 3.33 | 1.045 | 3.17 |
| Silicon | 7.04 | 1 | 7.04 |

**Table S9.** Eigenfrequency of the breathing mode for D2 for unscaled geometry and scaled geometry with indicated AMF.

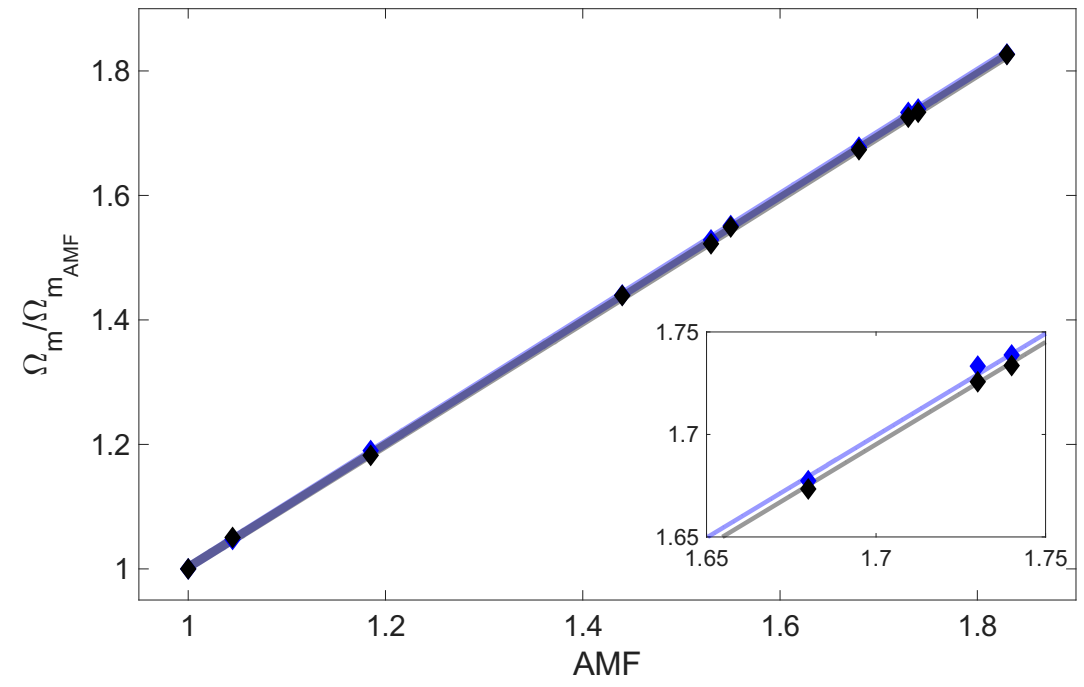


**Fig. S4.** Ratio of the mechanical breathing modes without and with AMF applied as a function of AMF.

## B. Optical and mechanical quality factors

While the value of the optical quality factor is affected by the defect region geometry, it is largely governed by the geometry of the mirror region, which is not a central topic of this work.

As already pointed out in the main text, the simulation of the mechanical quality factor yields unphysically large results, as dominant damping channels such as material and surface defects are not accounted for.

Both the optical and the mechanical quality factor have been simulated for silicon, diamond and 3C-SiC, see Tab. 1 of the main text. Given the above constraints, we do not repeat the simulations of the quality factors for the other materials discussed in this work.

## C. Optomechanical coupling

The vacuum optomechanical coupling rate $g_0$ receives contributions from the moving boundary and photoelastic effects [2], and can be written as

$$g_0 = (g_{mb} + g_{pe})\, x_{zpf}, \tag{S4}$$

where $g_{mb}$ and $g_{pe}$ are the moving boundary and photoelastic contributions per unit displacement, respectively, and $x_{zpf} = \sqrt{\hbar/(2m_{eff}\Omega_m)}$ is the amplitude of the zero-point fluctuations, with $\hbar$ the reduced Planck constant, $m_{eff}$ the effective mass, and $\Omega_m$ the eigenfrequency of the mechanical

breathing mode. The individual contributions are given by

$$g_{\mathrm{mb}} = -\frac{\omega_0}{2} \frac{\int_{\partial V} \left[\Delta\varepsilon \, |E_{\parallel}|^2 - \Delta(\varepsilon^{-1}) \, |D_{\perp}|^2\right] (\mathbf{u} \cdot \hat{\mathbf{n}}) \, dS}{\int_V \varepsilon \, |E|^2 \, dV}, \tag{S5}$$

$$g_{\mathrm{pe}} = -\frac{\omega_0}{2} \frac{\int_V \varepsilon^2 E_i^* \, p_{ijkl} \, S_{kl} \, E_j \, dV}{\int_V \varepsilon \, |E|^2 \, dV}, \tag{S6}$$

where $\omega_0 = 2\pi\nu$ is the optical angular eigenfrequency, $E$ ($E^*$) the electric field (and its complex conjugate) and $E_{\parallel}$ the electric field component parallel to the boundary, $D_{\perp}$ the electric displacement field component perpendicular to the boundary are obtained from the electrostatic finite element simulation. The normal displacement of the boundary $\mathbf{u} \cdot \hat{\mathbf{n}}$, and the stress tensor components $S_{kl}$, as well as the effective mass $m_{\mathrm{eff}}$, are taken from the elastic finite element simulations. The permittivity $\varepsilon = n^2$ (and $\Delta\varepsilon$ as well as $\Delta(\varepsilon^{-1})$ their (inverse) difference) is computed using the values in Tab. S1. The photoelastic tensor of the material $p_{ijkl}$ is not well documented for all material platforms discussed in this work. We therefore focus on those discussed in Tab. 1 of the main text. The photoelastic coefficients (in Voigt notation) for silicon, diamond and 3C-SiC used in these simulations are listed in Table S10.

| Material | $p_{11}$ | $p_{12}$ | $p_{44}$ |
|---|---|---|---|
| Silicon [3] | -0.1 | 0.01 | -0.05 |
| Diamond [4, 5] | -0.25 | 0.043 | -0.172 |
| 3C Silicon Carbide [6] | -0.11 | -0.03 | -0.01 |

**Table S10.** Photoelastic coefficients for calculating the optomechanical coupling for silicon, diamond and 3C-SiC.

## 5. EXPERIMENTAL SETUP

The experimental setup used to measure the optical cavity modes of OMCs as a function of the AMF discussed in Sec. 5 of the main text is depicted in Fig. S5. A widely tunable laser (1500 nm - 1630 nm) is fiber-coupled and passed through a polarizer and a circulator through which the light is led into a single-mode coupling fiber with a tapered edge, fabricated by etching in hydrofluoric acid [7, 8]. The fiber taper is aligned under visual inspection with the coupling waveguide fabricated alongside the OMC. It features a inverse taper which allows to evanescently couple light into the waveguide and from there into the defect region of the OMC. The coupling waveguide is terminated with a mirror to enable probing the cavity response via the reflection signal. The reflected light passes through the circulator and is detected using a photodetector. The laser wavelength is swept to record the response spectrum of the OMC. It exhibits sharp dips corresponding to the optical eigenmodes of the cavity, on top of a modulated background stemming from interferences in the coupling waveguide.

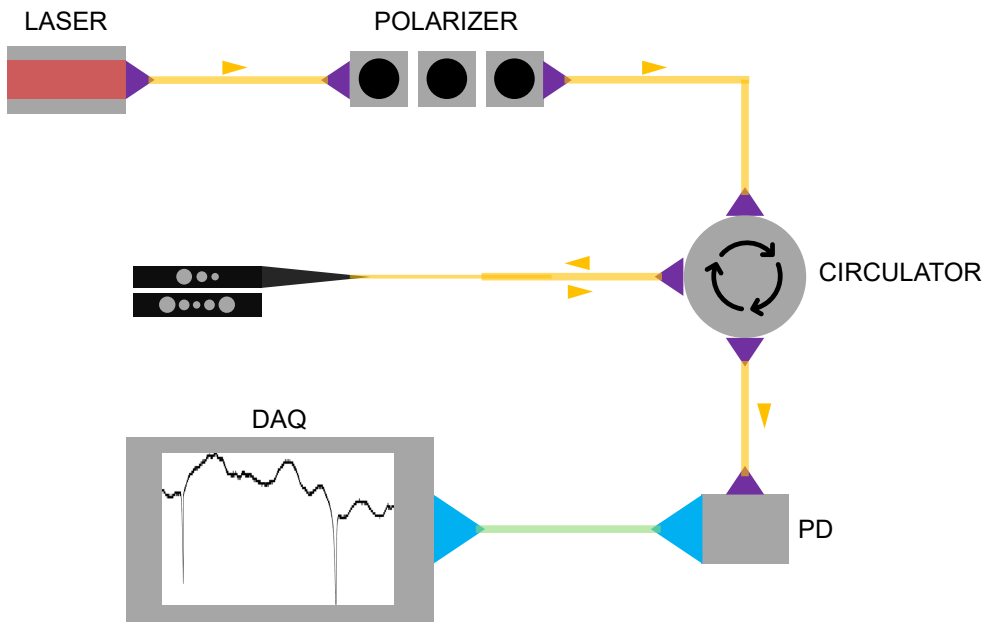


**Fig. S5.** Experimental measurement setup to characterize the optical eigenfrequencies of optomechanical crystals.



\end{document}